\documentclass[11pt,a4paper]{revtex4}
\usepackage{bm}
\usepackage{amsmath}
\usepackage{amsfonts}
\usepackage{hyperref}
\usepackage{graphicx}
\def\dd{\mathrm{d}}

\def\mcP{\mathcal{P}}
\def\mcR{\mathcal{R}}

\def\Mpl{M_{\rm Pl}}

\def\0{{(0)}}
\def\sig0{\dot{\sigma}_0}
\def\dsig{\delta \sigma}
\def\dsigd{\dot{\delta\sigma}}

\def\dphi{\delta \phi}
\def\ph0{\dot{\phi}_0}
\def\dphid{\dot{\delta \phi}}

\def\dn{\Delta n}

\begin{document}

\title{
Footprint of Two-Form Field:
Statistical Anisotropy in Primordial Gravitational Waves
}

\author{Ippei Obata}\thanks{obata@icrr.u-tokyo.ac.jp}
\affiliation{Institute for Cosmic Ray Research, The University of Tokyo, 5-1-5 Kashiwa-no-Ha, Kashiwa, Chiba, 277-8582, Japan}
\author{and Tomohiro Fujita}\thanks{t.fujita@tap.scphys.kyoto-u.ac.jp}
\affiliation{Department of Physics, Kyoto University, Kyoto, 606-8502, Japan}

\begin{abstract}
 We study the observational signatures of two-form field in the inflationary cosmology.
 In our setup a two-form field is kinetically coupled to a spectator scalar field and generates sizable gravitational waves and smaller curvature perturbation.
 We find that the sourced gravitational waves have a distinct signature:
they are always statistically anisotropic and their spherical moments are non-zero for hexadecapole and tetrahexacontapole, while the quadrupole moment vanishes.
 Since their amplitude can reach $\mathcal{O}(10^{-3})$ in
the tensor-to-scalar ratio, we expect this novel prediction will be tested
in the next generation of the CMB experiments.
\end{abstract}




\maketitle

%
%
%
\section{Introduction}

 The inflationary scenario elegantly explains the anisotropy of cosmic microwave background radiation (CMB) and the seed of the large scale structure in our universe.
 On top of them, it quantum-mechanically generates the fluctuations of spacetime, namely primordial gravitational waves, and imprints the B-mode polarization pattern in the CMB map.
 The detection of the primordial B-mode polarization originating from the inflationary universe is therefore one of the most important targets in cosmology.
 Its amplitude is parameterized by tensor-to-scalar ratio $r$ and recent joint collaboration of Planck and BICEP2/Keck array have constrained its amount as $r \lesssim 0.07$ \cite{Ade:2015lrj}.
 In the next decades, the sensitivity will increase up to $r \sim 10^{-3}$ by the appearance of LiteBIRD \cite{Matsumura:2013aja} and CMB-S4 \cite{Abazajian:2016yjj}.
 The energy scale probed by CMB observations is around the scale of grand unification theory $10^{16}\text{GeV}$, and thus we have a chance to obtain indispensable clues to develop the high energy physics such as GUT, supergravity or superstring through the detection of primordial gravitational waves.

 Conventionally primordial gravitational waves are considered to be provided by the vacuum fluctuation, whose power spectrum is almost scale invariant (slightly red-tilted) and statistically isotropic.
 However, these features are not necessarily true if the matter sector
significantly contributes to the generation of gravitational waves in the early universe.
 In a reduced four-dimensional action of string theory, for instance, a dilatonic scalar sector is generically coupled to an one-form field (gauge field) or a two-form field through their kinetic functions.
 Once these couplings are introduced during inflation, the time variation of kinetic function can amplify the quanta of the form field on superhorizon scales.
 Among these couplings, the particle production of U(1) gauge field has been motivated to explain the presence of intergalactic magnetic field \cite{Ratra:1991bn,Martin:2007ue, Demozzi:2009fu, Kanno:2009ei,Durrer:2010mq, Fujita:2012rb, Fujita:2013pgp, Fujita:2014sna, Obata:2014qba, Fujita:2016qab,Caprini:2017vnn}.
Furthermore, some models of inflation have been investigated in the framework of anisotropic inflation, where the inflaton is kinetically coupled to U(1) gauge field or two-form field~\cite{Watanabe:2009ct, Watanabe:2010fh, Kanno:2010nr, Watanabe:2010bu, Do:2011zza, Soda:2012zm, Bartolo:2012sd, Ohashi:2013qba, Ohashi:2013pca, Naruko:2014bxa, Abolhasani:2015cve, Ito:2016aai, Ito:2017bnn, Fujita:2017lfu, Ohashi:2013mka, Ito:2015sxj}.
 In these models the background form field naturally appears owing to the amplification on large scales and breaks the isotropy of universe.
The broken rotational invariance caused by the presence of background form field allows the perturbation of form field to interact with other scalar or tensor perturbations at linear level.
 As a result, the power spectra of some observables can be statistically anisotropic due to the enhanced perturbation of the form fields.
 The generation of such statistical anisotropy was originally motivated to explain the quadrupole anisotropy of the temperature fluctuation in the WMAP data \cite{2010ApJ...722..452G}, while current Planck data has not observed this signal and implies that its amplitude should be small, if any~\cite{Kim:2013gka, Rubtsov:2014yua, Ade:2015lrj, Ramazanov:2016gjl}.
 It is interesting to note that a little attention was paid to the statistical anisotropy of the primordial gravitational waves so far, because its generation by the U(1) gauge field is slow-roll suppressed compared to that of the curvature perturbation in the original model \cite{Watanabe:2010fh} and it is not produced at all by the two form field~\cite{Ohashi:2013mka}.

 Recently, however, it has been found that sizable amount of statistically anisotropic gravitational waves can be provided in an extended model of anisotropic inflation \cite{Fujita:2018zbr}.
 In this scenario, a U(1) gauge field is coupled to a spectator scalar field which enables to avoid the overproduction of statistical anisotropy in the curvature perturbation.
 Furthermore, the  mixing between the linear perturbations of the U(1) gauge field and the spectator field generates higher statistical anisotropies beyond quadrupole in the tensor power spectrum.
 This is a totally new prediction from the model of anisotropic inflation and incentivizes the observational search for the statistical anisotropy of the tensor perturbation.
 Hence, now it is time to revisit the case of two-form field and explore its new prediction in the extended scenario.
 In this work, we study a model of inflation where a two-form field kinetically coupled to a spectator scalar field.
 This situation allows the sizable mixing between the perturbations of the spectator scalar and the two-form field so that the amplified form field fluctuation sources that of the spectator field.
 Remarkably, we find that the sourced spectator field produces gravitational waves and finally generate statistically anisotropies in the tensor power spectrum.
 Intriguingly, the statistical anisotropies does not depend on the model parameters and higher harmonics beyond quadrupole moment are created.

 This paper is organized as follows.
In section \ref{Model Action and Setup}, we explain our model and explore the time evolution of the background fields.
In section \ref{Perturbation Dynamics}, we solve the linear perturbations
of the two-form field and the spectator scalar field. 
The productions of the curvature perturbation and gravitational waves,
in particular their statistical anisotropies, are studied in section \ref{Generation of Statistical Anisotropy}.
The detectability of the prediction of our model is discussed in section \ref{Detectability}.
Finally we present our conclusions in section \ref{Conclusion} with prospects for future work.

\section{Model Action and Background Dynamics}
\label{Model Action and Setup}

In this section, we present our model where a spectator scalar field is coupled to a  2-form field in the inflationary universe.
The Lagrangian density reads
\begin{equation}
\mathcal{L} = \dfrac{\Mpl^2}{2}R - \frac{1}{2}(\partial_\mu \phi)^2 -U(\phi) - \frac{1}{2}(\partial_\mu \sigma)^2 -V(\sigma)
- \frac{1}{12}I^2(\sigma)H_{\mu\nu\rho}H^{\mu\nu\rho},
\label{model action}
\end{equation}
where $R$ is the Ricci scalar, $\Mpl$ is the reduced Planck mass, $\phi$ is the inflaton, $\sigma$ is the spectator scalar field and $H_{\mu\nu\rho} = \partial_\mu B_{\nu\rho} + \partial_\nu B_{\rho\mu} + \partial_\rho B_{\mu\nu}$ is the field strength of two-form field $B_{\mu\nu}$.
$U(\phi)$ and $V(\sigma)$ are the potentials of these scalar fields.
The spectator scalar field $\sigma$ is coupled to the kinetic term of
the the form field via $I(\sigma)$.
We decompose these fields into the backgrounds and perturbations as
\begin{align}
&\phi(t,\bm{x})=\bar{\phi}(t)+\delta\phi(t,\bm{x}),\quad 
\sigma(t,\bm{x})=\bar{\sigma}(t)+\dsig(t,\bm{x}),\\
&B_{ij}(t,\bm{x}) = \bar{B}_{ij}(t) + \delta B_{ij}(t, \bm{x}) \ ,
\end{align}
where for the form field the gauge conditions $\bar{B}_{0i}(t) =\partial_i B_{ij}(t,\bm x) = 0,$ are taken.
 We present the gauge transformation of form field in Appendix \ref{Gauge transformation of form field}.
In the following discussion, we have eliminated $B_{0i}(t,\bm{x})=\delta B_{0i}(t,\bm{x})$ by solving the gauge constraint equations.
We approximate the background metric by the flat Robertson-Walker metric $ds^2 = -dt^2 + a(t)^2d\bm{x}^2$.
Note that although the background form field breaks the isotropy of the universe, we can correctly calculate the statistical anisotropy of perturbations even in this isotropic spacetime as far as the energy density of form field is subdominant.

Regarding the dynamics of the inflaton, we do not specify its potential form $U(\phi)$ and parameterize the cosmic expansion with a constant Hubble parameter, $H \simeq \text{const}.$ 
 On the other hand, for $V(\sigma)$ and $I(\sigma)$ we need to fix them for concrete calculations. 
 For the kinetic function $I(\sigma)$, we simply assume an exponential form
\begin{equation}
I(\sigma) = e^{\sigma/\Lambda} \ .
\end{equation}
Regarding the potential $V(\sigma)$,  we consider the same form as Ref.~\cite{Fujita:2018zbr}:
\begin{align}
V(\sigma)&= \mathcal{M}^3\frac{\sigma^2}{\sigma+\Lambda}
\ \sim\ \begin{cases}
\mathcal{M}^3\sigma\qquad \quad (\sigma\gg \Lambda)  \\
\mathcal{M}^3\sigma^2/\Lambda\ \quad (\sigma\ll \Lambda)  \\
\end{cases}.
\label{potential V}
\end{align}
In $V(\sigma)$ and $I(\sigma)$, we introduce two dimensionful parameters, $\Lambda$ and $\mathcal{M}$.
The above potential $V(\sigma)$ is well approximated by a linear potential for $\sigma \gg \Lambda$ where $\sigma$ slowly rolls down and by a quadratic potential for $\sigma\ll \Lambda$ where $\sigma$ gets stabilized by a significantly large potential curvature.
 Note that other forms of potential are also expected to provide similar dynamics
and predictions, if it implements the slow-roll and stabilization of $\bar{\sigma}$.

Let us study the dynamics of the background fields.
The model action eq.~\eqref{model action} leads to the following background equations:
\begin{equation}
\ddot{\bar \sigma}+3H\dot{\bar \sigma}+\bar{V}'= \frac{2}{\Lambda}\bar{\rho}_E,
\qquad
\dfrac{d}{dt}\left(\dfrac{1}{a}\bar{I}^2\dot{\bar{B}}_{ij}\right) = 0,
\label{BG EoM}
\end{equation}
with the energy density of the background form field,
\begin{equation}
\bar{\rho}_E = \frac{\bar{I}^2}{4a^4}\dot{\bar{B}}_{ij}^2 \equiv \frac{1}{4}\bar{E}_{ij}^2 \ .
\end{equation}
Here, $\bar{I}\equiv I(\bar{\sigma})$ is the background kinetic function, and dot and prime denote the cosmic time derivative and the derivatives with respect to fields (e.g., $\bar{V}'\equiv \partial_\sigma V(\bar{\sigma})$), respectively. 
The equation of motion (EoM) for $\bar{B}_{ij}$ can be integrated and one finds
$\bar{\rho}_E \propto a^{-2} \bar{I}^{-2}$ which is solely determined by $\bar{\sigma}(t)$.
 As we see below, the evolution of $\bar{\rho}_E$ is characterized by the following three phases. 
(i)  Growing phase:
Since its energy density is negligibly small in this phase, the contribution from the form field to the EoM of $\bar{\sigma}$  can be ignored, $|\bar{V}'|\gg 2\bar{\rho}_E/\Lambda$.
The slow-roll (terminal) velocity of $\bar{\sigma}$ is determined by $\bar{V}'$. 
Then the kinetic energy of $\bar{\sigma}$ is transferred to the form field and
$\bar{\rho}_E$ increases.
(ii) Attractor phase:
As $\bar{\rho}_E$ grows, the contribution from the form field to the EoM of $\bar \sigma$ becomes no longer negligible. Then the velocity of $\bar{\sigma}$ slows down and the decelerated evolution of the kinetic function makes $\bar{\rho}_E$ stay constant. (iii) Damping phase: When $\bar{\sigma}$ reaches $\Lambda$, it starts damped oscillations due to its quadratic potential. Since $\bar{I}$ practically stops evolving,  $\bar{\rho}_E$ decays as $a^{-2}$.

Approximate solutions for these three phases can be found from the EoMs as follows. In the slow-roll regime of $\sigma$ in which $\bar{\sigma}\gg\Lambda$, approximating $\bar{V}'\simeq \mathcal{M}^3$ and $\ddot{\sigma}\simeq 0$ in eq.~\eqref{BG EoM}, one finds the analytic solution of the EoM as
\begin{equation}
\bar{\sigma}(t)=\sigma_{\rm in}-\frac{\mathcal{M}^3}{3H}(t-t_{\rm in})+\frac{\Lambda}{2}\ln
\left[1+\frac{2\bar{\rho}_E(t_{\rm in})}{3\dn H^2\Lambda^2}
\left(\left(\frac{a}{a_{\rm in}}\right)^{2\dn}-1\right) \right],
\label{bs solution}
\end{equation}
where an exponentially decaying term is neglected, subscript ``in'' denotes the initial value, and we introduce an almost constant parameter $n$~defined as
\begin{equation}
n\equiv\frac{\mathcal{M}^3}{3H^2\Lambda},
\qquad
\dn\equiv n-1.
\end{equation}
Here we assume that $\bar{\rho}_E$ is set to be negligibly small at the initial time by some mechanisms.
For $\dn>0$, the term proportional to $\bar{\rho}_E(t_{\rm in})a^{2\dn}$,  which is initially negligible, eventually dominates the logarithm term in eq.~\eqref{bs solution} and it causes the shift from the growing phase into the attractor phase.
For $\bar{\sigma}\lesssim \Lambda$, however, the kinetic function stops evolving $\bar{I}\simeq 1$ and the effective mass of $\bar{\sigma}$ is given by
\begin{equation}
\bar{V}''\simeq \frac{2\mathcal{M}^3}{\Lambda} =  6n H^2 \qquad (\sigma\lesssim \Lambda).
\end{equation}
Therefore, assuming $n>1$ and $\bar{\rho}_E$ is initially small,  we find
the three phases of the background evolution,
\begin{align}
\dot{\bar \sigma}(t) &\simeq -H \Lambda\times
\begin{cases} n & (t<t_A) \\
1 & (t_A<t<t_D) \\
(a/a_D)^{-3/2}\cos(\sqrt{6n}H t +\delta) & (t_D<t) \\
\end{cases},
\label{s0 evolution}
\\
\bar{\rho}_E(t) &\simeq \frac{3}{2}\dn H^2 \Lambda^2\times
\begin{cases}  \,(a/a_A)^{2\dn} & (t<t_A) \\
1 & (t_A<t<t_D) \\
(a/a_D)^{-2} & (t_D<t) \\
\end{cases},
\label{rhoE evolution}
\end{align}
where $t_A$ and $t_D$ are the time when $\bar \rho_E$ reaches the attractor value $\frac{3}{2}\dn H^2 \Lambda^2$ and $\bar \sigma$ reaches $\Lambda$, respectively. We denote the values of the scale factor at these transition times by $a_A\equiv a(t_A)$ and $a_D\equiv a(t_D)$.
$\delta$ is a constant phase of the damped oscillation of $\bar \sigma$.
%
\begin{figure}[tbp]
    \hspace{-2mm}
  \includegraphics[width=70mm]{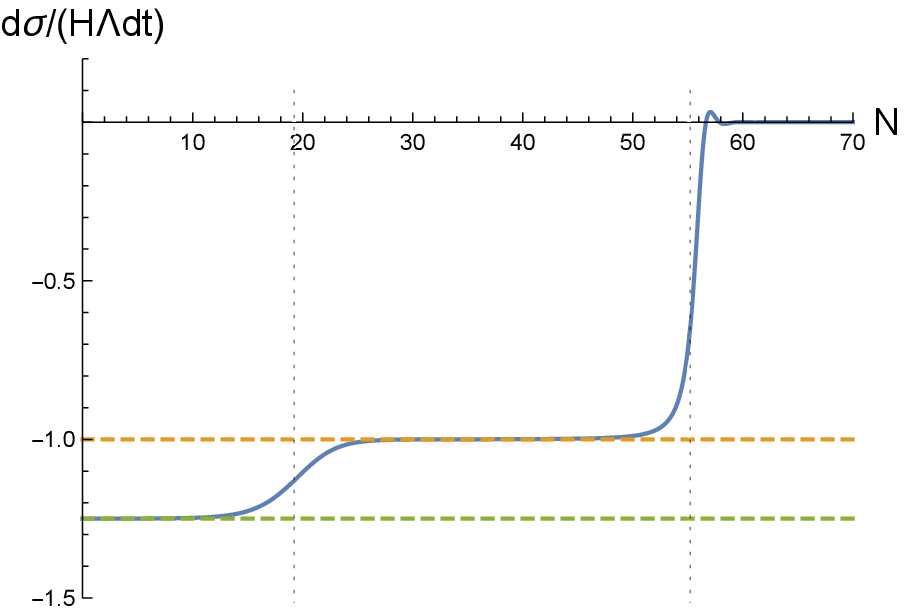}
  \hspace{5mm}
  \includegraphics[width=70mm]{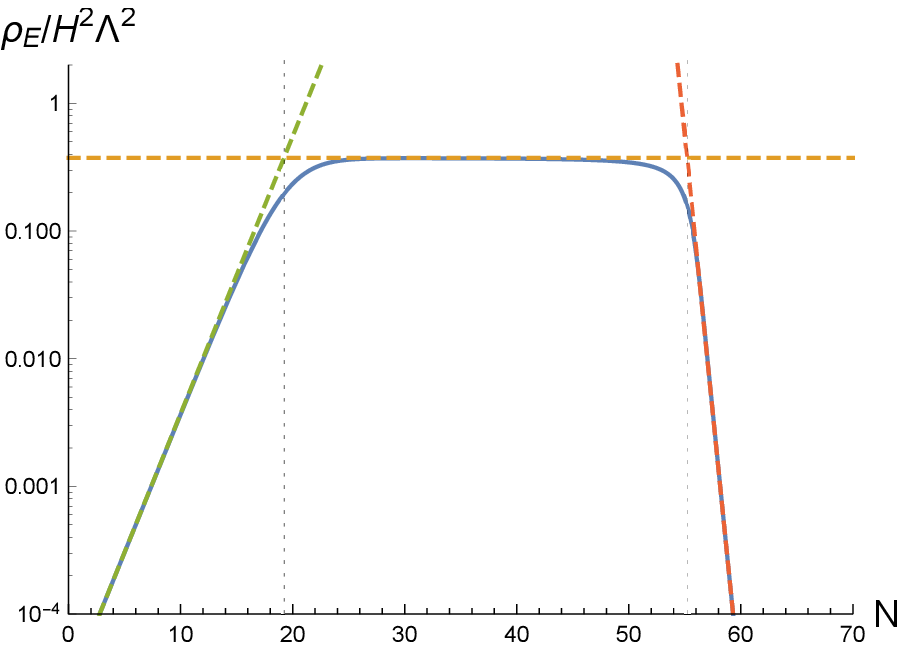}
  \caption
 {A numerical result of the time evolution of $\dot{\bar{\sigma}}$ (left panel) and $\bar{\rho}_E$ (right panel). The horizontal axis is e-folding number $N\equiv \ln (a/a_{\rm in})$. We set $n=1.25$ and the initial condition 
$\bar{\sigma}=60\Lambda, \dot{\bar{\sigma}}=-nH\Lambda, \bar{\rho}_E=2.5\times10^{-5}H^2\Lambda^2$
 at $N=0$.
The green dot-dashed lines represent the analytic solutions in the growing phase, $\dot{\bar \sigma}=-nH\Lambda$ (left panel) and $2.5\times10^{-5}H^2\Lambda^2\,a^{1/2}$ (right panel).
One can also see the analytic solutions in the attractor phase, $\dot{\bar \sigma}=-H\Lambda$ and $\bar{\rho}_E = \frac{3}{2}\dn H^2\Lambda^2$, 
which are shown as yellow dashed lines are realized.
The transition times between the phases are illustrated as the vertical black dashed lines.
 The red dashed line in the right panel indicates $\bar{\rho}_E$ decays as $a^{-2}$ in the damping phase.}
 \label{BG numerical}
\end{figure}
%
In figure~\ref{BG numerical}, we compare our analytic expressions with the numerical evaluation of $\bar{\sigma}(t)$ and $\bar{\rho}_E(t)$ for $n=1.25$, and they show excellent agreements.
Regarding $\Lambda$, we choose its value to satisfy $\Lambda \ll \Mpl$ so that the spectator energy density is subdominant.
 We will give a detailed discussion about the constraints on the background parameters in section \ref{Detectability}.

\section{Perturbation Dynamics}
\label{Perturbation Dynamics}

In this section, we discuss $\delta \sigma$ and $\delta B_{ij}$.
We quantize them, numerically solve their EoMs, and find approximate analytic solutions.
We mainly consider the modes which exit the horizon during the growing phase $(t<t_A)$, because the modes on smaller scales are never amplified
and it is harder for these modes to leave an observable imprint as we see in the next section.

\subsection{Quantization and numerical calculation}

We first decompose $\delta B_{ij}$ with an antisymmetric tensor $\epsilon_{ij}(\hat{\bm k})$ in Fourier space as
\begin{align}
\delta B_{ij}(t,\bm x) &= 
\int \frac{\dd^3 k}{(2\pi)^3} e^{i\bm{k}\cdot\bm{x}}\, \epsilon_{ij}(\hat{\bm{k}}) \delta B_{\bm k} \ .
\end{align}
 The antisymmetric tensor obeys the following relationships
\begin{equation}
k_i\epsilon_{ij}(\hat{\bm{k}}) = 0 \ , \quad \epsilon_{ij}(-\hat{\bm{k}}) = \epsilon^*_{ij}(\hat{\bm{k}}) \ , \quad \epsilon_{ij}(\hat{\bm{k}})\epsilon^*_{ij}(\hat{\bm{k}}) = 2 \ .
\end{equation}
 When we set the wave vector as $\hat{\bm{k}} = (\sin\theta\cos\varphi, \sin\theta\sin\varphi, \cos\theta)$, $\epsilon_{ij}$ is written as
\begin{equation}
\epsilon_{ij}(\hat{\bm{k}}) = i
\begin{pmatrix}
0 & \cos\theta & -\sin\theta\sin\varphi \\
-\cos\theta & 0 & \sin\theta\cos\varphi \\
\sin\theta\sin\varphi & -\sin\theta\cos\varphi & 0 \\
\end{pmatrix} \ .
\end{equation}
Without loss of generality, we can assume that the component of background form field is directed to the $(x,y)$ axes,
\begin{equation}
\bar{B}_{\mu\nu} = 
\begin{pmatrix}
0 & 0 & 0 & 0 \\
0 & 0 & B_{xy} & 0 \\
0 & -B_{xy} & 0 & 0 \\
0 & 0 & 0 & 0 \\
\end{pmatrix} \ .
\end{equation}
In that case, the inner product between the background form field and the antisymmetric tensor is
\begin{equation}
\sum_i\bar{E}_{ij} \,\epsilon_{ij}(\hat{\bm{k}}) = 2i\sqrt{2\bar{\rho}_E}\cos\theta \ .
\label{inner product}
\end{equation}
The Fourier transformations of $\delta\sigma$ is as usual,
\begin{equation}
\delta \sigma(t,\bm x)= 
\int \frac{\dd^3 k}{(2\pi)^3} e^{i\bm{k}\cdot\bm{x}}
\delta \sigma_{\bm k}(t).
\end{equation}
We calculate the quadratic action of $\delta\sigma_{\bm k}(\eta)$ and $\delta B_{\bm k}(\eta)$ in the spatially flat gauge where the non-dynamical scalar metric perturbations are integrated out.
 Neglecting slow-roll corrections and Planck suppressed terms, the resultant quadratic action is given by
\begin{equation}
S^{(2)}_\Delta=\frac{1}{2}\int\dd\eta\frac{\dd^3 k}{(2\pi)^3}\left[
\partial_\eta\Delta^\dag \partial_\eta\Delta+ \partial_\eta\Delta^\dag K \Delta- \Delta^\dag K \partial_\eta\Delta
-\Delta^\dag \Omega^2 \Delta\right],
\end{equation}
with 
\begin{align}
&\Delta
=\begin{pmatrix}a\dsig_{\bm k} \\
a^{-1}\bar{I}_B \delta B_{\bm k} \\
\end{pmatrix},
\quad
K=\dfrac{\sqrt{2\bar{\rho}_E}}{\Lambda H\eta}(i\cos\theta_{\bm
{k}})\begin{pmatrix}0 & 1 \\
1 & 0 \\
\end{pmatrix},
\notag\\
&\Omega^2 = \notag \\
& \begin{pmatrix}k^2-(2-\mu_\sigma^2/H^2)/\eta^2 & 
-\sqrt{2\bar{\rho}_E}(i\cos\theta) \partial_\eta(\ln[\bar{I}/a^2])/(\Lambda H \eta) \\
\sqrt{2\bar{\rho}_E}(i\cos\theta) \partial_\eta(\ln[\bar{I}/a^2])/(\Lambda H \eta) & k^2 - \partial_\eta^2 \bar{I}/\bar{I} + 2\partial_\eta a \partial_\eta \bar{I}/(a\bar{I}) \\
\end{pmatrix},
\end{align}
where $\eta$ is the conformal time and $\mu_\sigma^2= \bar{V}''- 4\Lambda^{-2}\bar{\rho}_E\cos(2\theta)$ (the full expression can be found in appendix~\ref{Quadratic Action}).
With these expressions, the EoMs are given by
\begin{equation}
\partial_\eta^2 \Delta+2K \partial_\eta \Delta+(\Omega^2 + \partial_\eta K)\Delta=0 \ .
\label{Matrix EoM}
\end{equation}
It should be noted that all the off-diagonal terms are proportional to $i\cos\theta$
with $\theta$  being the angle between $\bm k$ and the background form field $\dot{\bar{B}}_{ij}$ (see eq.~\eqref{inner product}).
 This is because the background form field breaks the isotropy of the universe which violates the decomposition theorem in perturbations and enable their couplings.
 When $\bm k$ is parallel to $\dot{\bar{B}}_{ij}$, this coupling disappears.

\begin{figure}[tbp]
    \hspace{-2mm}
  \includegraphics[width=70mm]{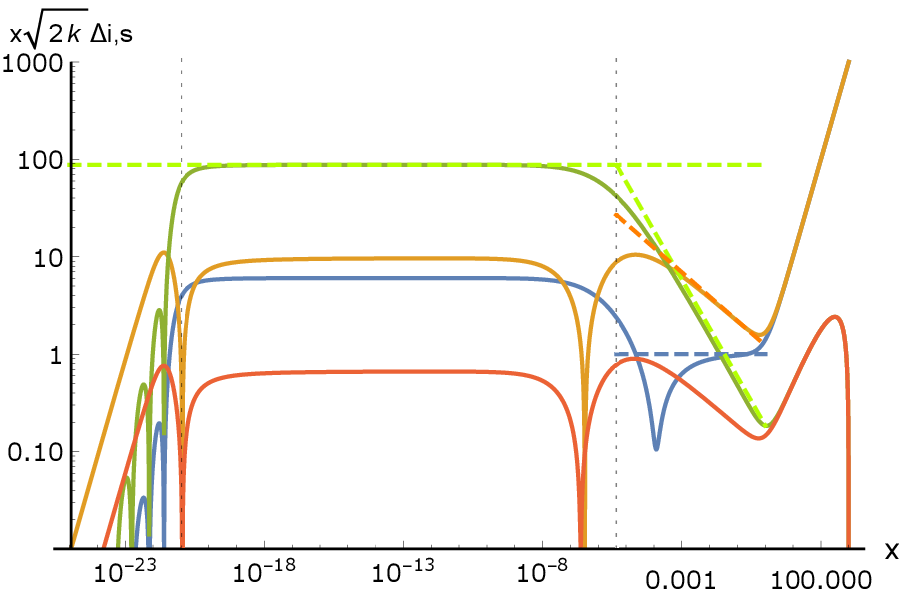}
  \hspace{5mm}
  \includegraphics[width=70mm]{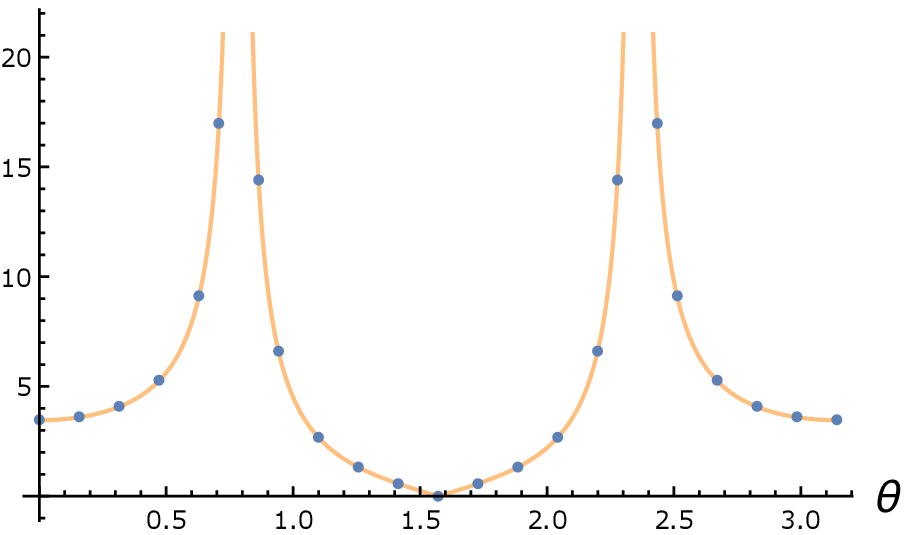}
  \caption
 {{\bf (Left panel)} The numerical results of $\sqrt{2k}x|a\delta\sigma^{\rm int}|$ (blue), $\sqrt{2k}x|\bar{I}\delta B^{\rm int}/a|$ (yellow), $\sqrt{2k}x|a\delta\sigma^{\rm src}|$ (green) and $\sqrt{2k}x|\bar{I}\delta B^{\rm src}/a|$ (red) are shown. The horizontal axis is $x\equiv -k\eta$. We fix $n=1.25$ and $\theta =\pi/5$. These modes exit the horizon before the background system enters the attractor phase at $x_A=4.4\times 10^{-6}$ and the damping phase at $x_D=1.1\times 10^{-21}$ (vertical black dashed lines). The other dashed lines in the figure are analytically derived in section~\ref{Analytic solutions}.
 {\bf (Right panel)} $|a\delta\sigma_k^p/(\bar{I}\delta B_k^q/a)|$ analytically derived in eq.~\eqref{super horizon ratio} (yellow line) and numerically obtained $|a\delta\sigma^{\rm src}_k/(\bar{I} \delta B^{\rm int}_k/a)|$ during the attractor phase (blue dots) are compared. The setting of the numerical calculation is the same as left figure. An excellent agreement is seen.}
 \label{PT numerical}
\end{figure}

 Since this system has both kinetic mixing and mass mixing, the coupled EoMs cannot be diagonalized and we have to solve the evolution of four modes which are the perturbations of $\dsig$ and $\delta B$ originating from the vacuum fluctuation of the respective fields. 
 Namely $\Delta$ should be promoted into mixed operators as
\begin{equation}
\hat{\Delta}
=\begin{pmatrix} a\delta\sigma^{\rm int}_k & a\delta\sigma^{\rm src}_k \\
\bar{I}\delta B_k^{\rm src}/a & \bar{I}\delta B_k^{\rm int}/a \\
\end{pmatrix}
\begin{pmatrix} \hat{a}_{\bm k} \\
\hat{b}_{\bm k} \\
\end{pmatrix}+{\rm h.c.} \ .
\label{Matrix quantization}
\end{equation}
The quantization is done by imposing the standard commutation relations
to two independent sets of annihilation/creation operators, $\{\hat{a}_{\bm k}, \hat{a}^\dag_{\bm k}\}$ and $\{\hat{b}_{\bm k}, \hat{b}^\dag_{\bm k}\}$.
The subscripts ``int'' and ``src'' represent the intrinsic modes and the sourced modes, respectively. 
Since $a\delta\sigma$ and $\bar{I}\delta B/a$ are decoupled in the sub-horizon limit, it is reasonable to assume that $a\delta\sigma_k^{\rm int}$ and $\bar{I}\delta B_k^{\rm int}/a$ are identical to the one for the  Bunch-Davies vacuum in the far past, while $a\delta\sigma_k^{\rm src}$ and $\bar{I}\delta B_k^{\rm src}/a$ vanish there:
\begin{equation}
\lim_{|k\eta|\to \infty}
\begin{pmatrix}a\delta\sigma_k^{\rm int}(\eta) & a\delta\sigma_k^{\rm src}(\eta) \\
\bar{I}\delta B_k^{\rm src}(\eta)/a & \bar{I}\delta B_k^{\rm int}(\eta)/a \\
\end{pmatrix}=\frac{e^{-ik\eta}}{\sqrt{2k}}\begin{pmatrix}1 & 0 \\
0 & 1 \\
\end{pmatrix}.
\end{equation}
 Here we introduce a dimensionless time variable $x\equiv -k\eta$.
The $x$ derivatives of the background scalar field $\bar{\sigma}(t)$ can be rewritten as
\begin{equation}
\partial_x\bar{\sigma}=-\frac{\dot{\bar \sigma}}{H x},
\qquad
\partial_x^2\bar{\sigma}=\frac{\ddot{\bar \sigma}+H\dot{\bar \sigma}}{H^2 x^2} \simeq \frac{\dot{\bar \sigma}}{H x^2}.
\end{equation}
 We numerically solve the above coupled EoMs, eq.~\eqref{Matrix EoM},
for modes that exit the horizon during the growing phase.
In figure~\ref{PT numerical}, we show the numerical results.
%
%
In the next subsection, we develop an analytic treatment to understand these numerical results.

\subsection{Analytic solutions}
\label{Analytic solutions}

The EoMs of mode functions are given by
\begin{align}
\left[\partial_x^2+1-\frac{2-\mu_\sigma^2/H^2}{x^2}\right](a\delta\sigma_k^p)
&=2\sqrt{2}(i\cos\theta)\frac{\sqrt{\bar{\rho}_E}}{H\Lambda\, x}\,\dfrac{\bar{I}}{a}\partial_x \delta B_k^q
,
\\
\left[\partial_x^2 +1-\frac{\partial_x^2 \bar{I}}{\bar{I}} - \frac{2\partial_x \bar{I}}{x\bar{I}}\right](\bar{I}\delta B_k^p/a)  &=
-2\sqrt{2}(i\cos\theta)\frac{\sqrt{\bar{\rho}_E}}{H\Lambda\,x}\, a\partial_x\delta\sigma_k^q,
\label{dA EoM}
\end{align}
where the superscripts take $(p,q)=$(int, src) or (src, int) and thus we have four equations.
$\bar{V}''$ in $\mu_\sigma^2 \equiv \bar{V}''- 4\Lambda^{-2}\bar{\rho}_E\cos(2\theta)$ can be ignored during the growing and attractor phases.
Then, although $\mu_\sigma^2$ is negative for $0<\theta<\pi/4, \ 3\pi/4<\theta<\pi$,
it does not lead to tachyonic instability  as we see soon.

\subsubsection{Growing phase}

During the growing phase, since $\bar{\rho}_E\ll H^2\Lambda^2$, all the terms with $\bar{\rho}_E$ including the coupling terms between $\delta\sigma$ and $\delta B$ are sub-leading.
Then it is straightforward to obtain the homogeneous solutions in the super-horizon limit as,
\begin{equation}
a\delta\sigma^{\rm int}_k
\simeq
\frac{i }{\sqrt{2k}\,x}, \qquad \dfrac{\bar{I} \delta B^{\rm int}_k}{a} \simeq \frac{\Gamma(n+\frac{1}{2})}{\sqrt{2\pi k}}\left(\frac{x}{2}\right)^{-n},
\qquad  (x\ll 1).
\label{ds before attractor}
\end{equation}
They are plotted as the blue and yellow dashed lines in the left panel of figure~\ref{PT numerical}. Note that $\bar{I} \delta B^{\rm int}_k/a^2$ becomes much larger than $\delta \sigma^{\rm int}_k$, because the former grows on super-horizon scales in proportion to $a^{\Delta n}$, while the latter stays constant.
We do not discuss $\bar{I}\delta B^{\rm src}_k/a$, which is sourced by $a\delta\sigma^{\rm int}_k$ and hence sub-leading (see the red line in figure~\ref{PT numerical}).

$a \delta\sigma^{\rm src}_k$ sourced by  $\bar{I}\delta B_k^{\rm int}/a$ on super-horizon scales during the growing phase can be obtained with the Green's function method. $\delta\sigma^{\rm src}_k$ can be calculated as
\begin{align}
a\delta\sigma^{\rm src}_k(x)
&= 2\sqrt{2}(i\cos\theta)\int\dd y\, G_{R}(x,y) \,  \frac{\sqrt{\bar{\rho}_E(y)}}{H\Lambda\, y}\,\dfrac{\bar{I}}{a}\partial_y \delta B^{\rm int}_k(y) \ , \label{G function} \\
G_{R}(x,y) &\equiv -\Theta(y-x)\,(x^3-y^3)/(3xy) \ .
\end{align}
 The retarded Green's function $G_{R}$ satisfies $\left[\partial_x^2-2/x^2\right]G_R (x,y)=\delta(x-y)$
in which the gradient term and the  mass term $\mu_\sigma^2$ are ignored.
 Defining $x_A$ as the time when $\bar{\rho}_E$ reaches the attractor value and integrating eq.~\eqref{G function}, we obtain
\begin{equation}
a\delta\sigma^{\rm src}_k\simeq \frac{1}{\sqrt{2k}x}\, \frac{2^{n+1}\Gamma(n+\tfrac{3}{2})}{\sqrt{3\pi \dn}}
(i\cos\theta)\, \frac{x_A^{\dn}}{x^{2\dn}}.
\qquad  (x\ll 1)
\label{sigma src}
\end{equation}
It is plotted in the left panel of figure~\ref{PT numerical} as a green dashed line.
Therefore $\delta \sigma^{\rm src}$ grows as $a^{2\Delta n}$ on super-horizon scales during the growing phase, faster than $\bar{I}\delta B^{\rm int}/a^2$.

\subsubsection{Attractor phase}

We can derive a simple relationship between $a\delta\sigma_k$ and $\bar{I}\delta B_k/a$ on super-horizon scales during the attractor phase.
Changing the time variable from conformal time to cosmic time, one can rewrite
the EoM of $\delta\sigma$ as 
\begin{align}
&\ddot{\delta\sigma}_k^p+3H\dot{\delta\sigma}_k^p-6\Delta n\cos(2\theta)H^2\delta\sigma_k^p
=2
\sqrt{3\Delta n}(i\cos\theta)\ H\frac{\bar{I}\delta\dot{B}_k^q}{a^2} \ , \label{att sig EoM} \\
&\partial_t\left(\dfrac{\bar{I}\delta\dot{B}_k^p}{a^2}\right)  = 2\sqrt{3\Delta n}(i\cos\theta)\delta\dot{\sigma}_k^q \ ,
\end{align}
where the spatial gradient terms are ignored and some background time dependence during the attractor phase is used.
 Then we can find that $\delta\sigma_k$ and $\bar{I}\delta B_k/a^2$ have a constant solution while the other solutions are decaying.
Focusing on the constant solution ($\delta\sigma=\text{const}. , \bar{I}\delta B_k\propto a^2$), we will find the following simple relation
which depends only on $n$ and $\theta$ :
\begin{equation}
\frac{a\delta\sigma_k^p}{\bar{I} \delta B_k^q/a}=
-\sqrt{\frac{3}{\Delta n}}\frac{i\cos\theta}{\cos 2\theta},
\qquad {\rm (super\ horizon)} \ .
\label{super horizon ratio}
\end{equation}
 This equation holds for both $(p,q)=$(int, src) and (src, int).
In the right panel of figure \ref{PT numerical} we show the case of $(p,q)=$(src, int) and confirm that this is indeed a good approximation of numerical results.
Now one needs to connect the solutions during the attractor phase
to the one during the growing phase to determine the amplitude
of $a\delta\sigma_k$ or $\bar{I}\delta B_k/a$. 
 As an approximation, we extrapolate $\delta\sigma^{\rm src}_k$ of the growing phase till the transition time $x=x_A\equiv -k\eta_A$.
 Substituting $x=x_A$ into eq.~\eqref{sigma src}, we obtain
\begin{equation}
\delta\sigma^{\rm src}_k = \frac{H}{\sqrt{2k}k}\, \gamma(n) (i\cos\theta)\, \left(\frac{k_A}{k}\right)^{\dn} \ , \qquad \gamma(n) \equiv \frac{2^{n+1}\Gamma(n+\tfrac{3}{2})}{\sqrt{3\pi \dn}}
\label{analytic dsig}
\end{equation}
where we rewrite $x_A^{-\Delta n}=\left(k_A/k\right)^{\dn}$ and $k_A$ is the wave number which exits horizon when the background enters the attractor phase.
 This expression is plotted in the left panel of figure~\ref{PT numerical} as a dark green dot-dashed line and we can see that \eqref{analytic dsig} is actually a good approximation on the constant evolution of $\delta\sigma^{\rm src}$ during the attractor phase.
 Using the relation~\eqref{super horizon ratio},
we also obtain the constant amplitude of $\bar{I}\delta B^{\rm int}/a^2$ as well.
 Both $\delta\sigma^{\rm src}_k$ and $\delta B^{\rm int}_k$ have red-tilted spectrum for $k<-\eta_A^{-1}$,
because they continue to grow from the horizon exit until the attractor phase starts.

\subsubsection{Damping phase}

 Since a perturbation on super-horizon scales behaves in the same way as its background component, $\delta \sigma_k$  oscillates with an amplitude decaying as $a^{-3/2}$ and $\bar{I} \delta \dot{B}_k/a^2$ decays as $a^{-1}$, which are indeed confirmed in the left panel of figure~\ref{PT numerical}.
 In the following sections, we calculate the generation of $\delta\phi$ and $h_{ij}$
by focusing on the attractor phase.

\section{Generation of Statistical Anisotropy}
\label{Generation of Statistical Anisotropy}

 In this section, we present the generation of statistical anisotropies in both inflaton and tensor perturbations.

\subsection{Sourced inflaton perturbation}

 Although the inflaton has no direct coupling to the spectator scalar and the two-form field, their linear perturbations are coupled via the gravitational coupling.
 The EoM for $\delta\phi(t,\bm x)$ is given by
\begin{equation}
\left[\partial_t^2+3H\partial_t-\frac{\nabla^2}{a^2}+\mu_\phi^2\right]\delta\phi
=-\Omega_{\phi\sigma}\delta\sigma-\Omega_{ij}^{B\phi}\frac{\bar{I}^2}{a^4}\delta B_{ij},
\label{dphi EoM}
\end{equation}
where the full expressions for $\mu_\phi^2,\ \Omega_{\phi\sigma}$ and $\Omega_{ij}^{B\phi}$
can be found in appendix~\ref{Quadratic Action}.
Since we are interested in a super-horizon mode sourced by ~$\delta\sigma$ ~and ~$\delta B_{ij}$ ~during attractor phase, eq.~\eqref{dphi EoM} can be reduced into
\begin{equation}
\left[\partial_x^2-\frac{2}{x^2}\right](a\delta\phi_k)
=-\frac{\Omega_{\phi\sigma}}{x^2H^2}\,a\delta\sigma^{\rm src}_k,
\end{equation}
where we have ignored the gradient term, the inflaton mass and the contribution
from the gauge field, because $\Omega^{B\phi}_{ij}$ is suppressed by slow-roll parameters compared to $\Omega_{\phi\sigma}$ 
while $a\delta\sigma^{\rm src}$ and $\bar{I}\delta B_k^{\rm int}/a$
are the same order due to the relation eq.~\eqref{super horizon ratio}.
We also used $\delta \sigma_k^{\rm src}\gg \delta \sigma_k^{\rm int}$, since we are interested in the perturbations on scales where $\delta\sigma^{\rm src}_k$ is amplified significantly during the growing phase (see figure~\ref{PT numerical}).
During the attractor phase, the coupling between $\delta\phi$ and $\delta\sigma$ is rewritten as
\begin{equation}
\Omega_{\phi\sigma} \simeq
-\frac{\dot{\bar \phi}\dot{\bar \sigma}}{\Mpl^2}\left[3-\frac{2\bar{\rho}_E\cos^2\theta}{\Lambda H\dot{\bar \sigma}}\right] = 3n\sqrt{2\epsilon_\phi}\,H^2\frac{\Lambda}{\Mpl}\left(1-\dfrac{\dn}{n}\sin^2\theta\right), 
\end{equation}
where $\dot{\bar \phi}/\Mpl H\simeq \sqrt{2\epsilon_\phi}$ is used.
Assuming $\epsilon_\phi \simeq \text{const}.$, we obtain the sourced inflaton perturbation as
\begin{align}
a\delta \phi^{(s)}= -\frac{\Omega_{\phi\sigma}}{H^2}\,\delta\sigma^{\rm src}\int \dd y \, G_R(x,y) \frac{a}{y^2}
= -\frac{\Omega_{\phi\sigma}}{3H^2}\,a\delta\sigma^{\rm src}(N_A-1/3),
\end{align}
where we have performed the time integration over only the attractor phase and $N_A$ denotes the e-fold number of the duration of the attractor phase.
Putting all together and dropping an overall minus sign, we find
\begin{equation}
\frac{\delta\phi^{(s)}}{\delta\phi^{\rm (vac)}}
=n\gamma(n)(i\cos\theta)\left(1-\dfrac{\dn}{n}\sin^2\theta\right) \sqrt{2\epsilon_\phi} \frac{\Lambda}{\Mpl}
\left(\frac{k_A}{k}\right)^{\dn}(N_A-1/3),
\end{equation}
where the amplitude of the vacuum contribution is $\delta\phi^{\rm (vac)}=H/\sqrt{2k^3}$.
Thus, as anticipated, the sourced $\delta\phi$ is suppressed by the slow-roll parameter $\epsilon_\phi^{1/2}$
and $\Lambda/\Mpl$, while it is boosted by $\left(k_A/k\right)^{\dn}$ and $N_A$ compared to the conventional vacuum fluctuation.
 Defining the dimensionless power spectrum $(2\pi)^3\delta(\bm{k}+\bm{k}')\mathcal{P}_\zeta(k) = \tfrac{k^3}{2\pi^2}\langle\zeta_{\bm{k}} \zeta_{\bm{k}'} \rangle$, the power spectrum of the sourced curvature perturbation for $k\ll k_A$ is
\begin{align}
\mcP_\zeta^{(s)}
&= \mcP_\zeta^{\rm (vac)}
\left[n\gamma(n)\, \sqrt{2\epsilon_\phi} \frac{\Lambda}{\Mpl}
\left(\frac{k_A}{k}\right)^{\dn}(N_A-1/3)\right]^2\cos^2\theta\left(1-\dfrac{\dn}{n}\sin^2\theta\right)^2 \ ,
\label{sourced Pz}
\end{align}
where $\mcP_\zeta^{\rm (vac)}\equiv H^2/(8\pi^2\Mpl^2 \epsilon_\phi)$,
which is the power spectrum of the curvature perturbation contributed only by the vacuum fluctuation of $\delta\phi$
as $\zeta_{\bm k}=-\delta\phi_{\bm k}/(\sqrt{2\epsilon_\phi}\Mpl)$.

\subsection{Sourced gravitational waves}

 We discuss the generation of gravitational waves in our model.
 The tensor perturbation $h_{ij}$ is defined as the fluctuation of the spatial component in metric $g_{ij} = a(t)^2(\delta_{ij} + h_{ij})$ which obeys the transverse and traceless conditions $\partial_ih_{ij} = h_{ii} = 0$.
 The quadratic action of gravitational waves is given in \eqref{eq: GWq} in appendix~\ref{Quadratic Action}.
 We decompose tensor perturbations into their Fourier modes
\begin{equation}
h_{ij}(t,\bm x)= 
\int \frac{\dd^3 k}{(2\pi)^3} e^{i\bm{k}\cdot\bm{x}}\, \left[e^+_{ij}(\hat{\bm k}) h_{\bm k}^+(t)+i e^\times_{ij}(\hat{\bm k}) h_{\bm k}^\times (t)\right] \ ,
\end{equation}
where
\begin{align}
e^+_{ij}(\hat{\bm k})&=\dfrac{1}{\sqrt{2}}\begin{pmatrix}
\cos^2\theta \cos^2\varphi-\sin^2\varphi\ & (\cos^2\theta+1)\sin\varphi\cos\varphi\ & -\sin\theta\cos\theta\cos\varphi \\
(\cos^2\theta+1)\sin\varphi\cos\varphi\ & \cos^2\theta\sin^2\varphi - \cos^2\varphi\ & -\sin\theta\cos\theta\sin\varphi \\
-\sin\theta\cos\theta\cos\varphi\ & -\sin\theta\cos\theta\sin\varphi & \sin^2\theta \\
\end{pmatrix} \ , \\
e^\times_{ij}(\hat{\bm k})&=\dfrac{1}{\sqrt{2}}\begin{pmatrix}
-2\cos\theta\sin\varphi\cos\varphi\ & \cos\theta(\cos^2\varphi - \sin^2\varphi)\ & \sin\theta\sin\varphi \\
\cos\theta(\cos^2\varphi - \sin^2\varphi)\ & 2\cos\theta\sin\varphi\cos\varphi\ & -\sin\theta\cos\varphi \\
\sin\theta\sin\varphi & -\sin\theta\cos\varphi\ & 0 \\
\end{pmatrix}
\end{align}
are polarization tensors satisfying the normalization and orthogonal conditions.
 Then, one can find that the interaction between $\delta B_{ij}$ and $h_{ij}$ in \eqref{eq: GWq} vanishes.
 This result is consistent with the previous study \cite{Ohashi:2013qba}.
 However, as is discussed in section \ref{Perturbation Dynamics}, not only $\delta B_{ij}$ but that of spectator field $\delta\sigma$ is also amplified and can contribute the generation of gravitational waves in our model.
 Neglecting slow-roll corrections, we find the EoMs for the sourced tensor mode functions, $h^+_k(t)$ and $h^\times_k(t)$, as
\begin{align}
\left[\partial_t^2+3H\partial_t+\frac{k^2}{a^2}\right]h^+_k &\simeq
\frac{4\sqrt{2}\bar{\rho}_E}{\Mpl^2} \frac{\dsig_k}{\Lambda} \sin^2\theta \ ,
\label{h+ EoM}
\\
\left[\partial_t^2+3H\partial_t+\frac{k^2}{a^2}\right]h^\times_k &\simeq
0 \ ,
\end{align}
where we have used the background equations during the attractor phase.
It is interesting to note that $h^+_k$ is sourced by $\delta \sigma_k$, while $h^\times_k$ is not.
 This is because $h^+_k$ only couples the scalar degree of freedom in this decomposition.
Introducing the canonical field,
\begin{equation}
\psi_{k}^\lambda \equiv \frac{1}{2}a\Mpl h_{k}^\lambda,
\qquad (\lambda=+,\times)
\end{equation}
and changing the time variable from the cosmic time to $x\equiv -k\eta$, one rewrites eq.~\eqref{h+ EoM} in the super-horizon limit as
\begin{align}
\left[ \partial_x^2-\frac{2}{x^2}\right]\psi^+
=\frac{3\sqrt{2}\dn}{x^2}\frac{\Lambda}{\Mpl}
\sin^2\theta \, a \delta\sigma^{\rm src},
\end{align}
where we used \eqref{rhoE evolution}. 
With the Green's function method, we obtain
\begin{equation}
\psi^{+}_{(s)}=i\frac{aH}{\sqrt{2k}k} \sqrt{2}\dn\gamma(n)\,
\cos\theta \sin^2\theta
\frac{\Lambda}{\Mpl}\left(\frac{k_A}{k}\right)^{\dn}(N_A-1/3).
\end{equation}
Thus, dropping the overall minus sign, we find that the sourced tensor perturbation divided by its vacuum fluctuation is given by
\begin{equation}
\frac{\psi^{+}_{(s)}}{\psi_{\rm (vac)}}=
i\sqrt{2}\dn\,\gamma(n)\,
\cos\theta \sin^2\theta
\frac{\Lambda}{\Mpl}\left(\frac{k_A}{k}\right)^{\dn}(N_A-1/3),
\end{equation}
where $\psi_{\rm (vac)}=aH/\sqrt{2k^3}$.

 The power spectrum of the sourced tensor perturbation for $k\ll k_A$ is therefore
\begin{align}
\mcP_h^{(s)}&=\frac{1}{2}\mcP_h^{\rm (vac)}\left|\frac{\psi^{+}_{(s)}}{\psi_{\rm (vac)}}\right|^2 = \frac{2H^2}{\pi^2\Mpl^2}g_*\left( \cos^2\theta -2\cos^4\theta +\cos^6\theta\right)  \ , 
\notag \\
g_* &\equiv \left[ \dn\,\gamma(n)\frac{\Lambda}{\Mpl}\left(\frac{k_A}{k}\right)^{\dn}\left(N_A-\frac{1}{3}\right) \right]^2
,
\label{sourced Ph}
\end{align}
where $\mcP_h^{\rm (vac)}=2H^2/(\pi^2 \Mpl^2)$.
 Remarkably, the angular pattern in $\mcP_h^{(s)}$ does not depend on any model parameters.
 Here we are interested in the statistical anisotropy of gravitational waves and analyze it with the spherical harmonics $Y_{lm}(\hat{\bm k})$.
The tensor power spectrum is expanded as
\begin{equation}
\mcP_h(\bm{k}) = \mcP_h^{\rm (iso)}(k)\left(1 + \sum_{l={\rm even}}^{\infty} \sum_{M=-l}^{l} g_{lM}Y_{lM}(\hat{\bm{k}})\right) \ ,
\end{equation}
where $l=2,4,6,...$ and the coefficients of $Y_{lm}$ are called 
the quadrupole moment $(l=2)$, the hexadecapole moment $(l=4)$, 
the tetrahexacontapole moment $(l=6)$, and so on.
Rewriting $\cos^n\theta$ into the the combinations of Legendre polynomial $P_n(\cos\theta)$
\begin{align}
\cos^2\theta &= \dfrac{1}{3} + \dfrac{2}{3}P_2(\cos\theta) \ , \\
\cos^4\theta &= \dfrac{1}{5} + \dfrac{4}{7}P_2(\cos\theta) + \dfrac{8}{35}P_4(\cos\theta) \ , \\
\cos^6\theta &= \dfrac{1}{7} + \dfrac{10}{21}P_2(\cos\theta) + \dfrac{24}{77}P_4(\cos\theta) + \dfrac{16}{231}P_6(\cos\theta)
\end{align}
and using the following relation
\begin{equation}
P_l(\cos\theta) = \dfrac{4\pi}{2l + 1}\sum_{M=-l}^lY^*_{lM}(\hat{\bar{\bm{B}}})Y_{lM}(\hat{\bm{k}}) \ ,
\end{equation}
one can find that the coefficient $g_{2M}$ reads
\begin{equation}
g_{2M} = \dfrac{g_*}{1+\tfrac{8}{105}g_*}\left(\dfrac{10}{21} - \dfrac{8}{7} + \dfrac{2}{3}\right)\dfrac{4\pi}{5}Y^*_{2M}(\hat{\bar{\bm{B}}}) = 0 \ .
\end{equation}
 Intriguingly the quadrupole moment $g_{2M}$ in the anisotropic tensor power spectrum exactly vanishes. Therefore only the hexadecapole and tetrahexacontapole
moments are non-zero. This particular property may be used as a smoking gun
of the existence of two-form field during inflation.

\section{Detectability}
\label{Detectability}

In this section, we explore the possibility that the gravitational waves produced in our model will be detected by upcoming CMB observations.
For the sourced gravitational waves to be detectable, it should be larger than the conventional inflationary ones from the tensor vacuum fluctuation.
At the same time, the curvature perturbation induced by $\delta\sigma$ should not exceed the contribution from the inflaton perturbation $\delta\phi$, because the former is red-tilted too much for $\Delta n = \mathcal{O}(1)$.
Thus we require the following two conditions.
\begin{equation}
\mathcal{R}_\zeta \equiv \mathcal{P}^{(\text{s})}_\zeta/\mathcal{P}^{(\text{vac})}_\zeta \ll 1 \ , \qquad \mathcal{R}_h \equiv \mathcal{P}^{(\text{s})}_h/\mathcal{P}^{(\text{vac})}_h \gtrsim 1.
\end{equation}
To satisfy these conditions, their ratio needs to be much larger than unity,
\begin{equation}
\dfrac{\mathcal{R}_h}{\mathcal{R}_\zeta} = \dfrac{8\Delta n^2}{n^2r_{\text{vac}}}\dfrac{(1-\cos^2\theta)^2}{\left(1-\tfrac{\dn}{n}\sin^2\theta\right)^2} \gg 1.
\end{equation}
where $r_{\rm vac}\equiv \mcP_h^{\rm (vac)}/\mcP_\zeta^{\rm (vac)}=16\epsilon_\phi$ is used. It is not difficult to find a set of the parameters satisfying this condition.
For instance, we find the parameters
\begin{equation}
n=1.25,\quad r_{\rm vac}=5\times 10^{-4},\quad
\Lambda=10^{-2}\Mpl,\quad k_{\text{CMB}}=e^{-10}k_A,
\quad N_A=30,
\label{ex para}
\end{equation}
with which $8\Delta n^2/(n^2r_{\text{vac}}) = 640$. In this case, the primordial gravitational waves are significantly enhanced, while the induced curvature perturbation is negligible,
\begin{align}
\mcR_h &\simeq 
20\left(\cos^2\theta -2\cos^4\theta +\cos^6\theta\right) \ ,
\\
\mcR_\zeta &\simeq
0.02\left(\cos^2\theta+0.5\cos^4\theta+0.0625\cos^6\theta\right) \ .
\end{align}
Although the tensor mode is apparently $\mathcal{O}(10)$ times amplified, one should notice that the angular dependence suppresses it.
The averaged values of the angular factors are
\begin{align}
\frac{1}{\pi}\int_0^{\pi}\dd\theta\,\left( \cos^2\theta -2\cos^4\theta +\cos^6\theta\right)&=\frac{1}{16} \ ,
\\
\frac{1}{\pi}\int_0^{\pi}\dd\theta\,\left(\cos^2\theta+0.5\cos^4\theta+0.0625\cos^6\theta\right)&\simeq0.71 \ .
\end{align}
Therefore, the tensor-to-scalar ratio is enhanced by an $\mathcal{O}(1)$ factor in this case,
\begin{equation}
\bar{r}_{\rm src} \equiv \frac{r_{\rm vac}}{\pi}\int_0^{\pi}\dd\theta \, \mcR_h(\theta) \simeq 6.3 \times 10^{-4}
\quad\Longrightarrow\quad
r=r_{\rm vac}+\bar{r}_{\rm src} \sim 10^{-3} \ .
\end{equation}
Since the upcoming CMB B-mode observations (e.g. LiteBIRD or CMB-S4) aim to achieve the sensitivity $r= \mathcal{O}(10^{-3})$, this enhanced primordial gravitational waves are potentially detectable with future CMB missions such as LiteBIRD \cite{Matsumura:2013aja} and CMB-S4 \cite{Abazajian:2016yjj}.

 Before closing this section, we discuss three constraints on the background dynamics in this model.
 First, we introduce the e-folding number
\begin{equation}
N_G\equiv \ln[k_A/k]
\end{equation}
from the horizon exit till the onset of the attractor phase, or the duration of the growing phase which the mode experiences,
\begin{equation}
\bar{\rho}_E(t_k) \exp[2\dn N_G]=  \frac{3}{2}\dn H^2 \Lambda^2
\quad\Longrightarrow\quad
N_G=\frac{1}{2\dn}\ln\left[\frac{3\dn H^2\Lambda^2}{2\bar{\rho}_E(t_k)}\right],
\end{equation}
where $t_k$ denotes the time when the $k$-mode of interest exits the horizon.
 We put an upper bound on $N_G$. 
As $\bar{\rho}_E(t_k)$ is smaller, $N_G$ becomes larger.
However, for the validity of the perturbative approach $\bar{B}_{ij} \gg \delta B_{ij}$, $\bar{\rho}_E(t_k)$ should be much larger than $\mathcal{O}(H^4)$.
Requiring $\bar{\rho}_E(t_k)>10^2H^4$ and eliminating $H$
with $r_{\rm vac}=2H^2/(\pi^2\Mpl^2 \mcP_\zeta^{\rm obs})$,
we obtain the upper bound on $N_G$ as
\begin{equation}
N_G < N_G^{\max}\equiv\frac{1}{2\dn}
\ln\left[\frac{3\dn }{10^2\pi^2r_{\rm vac}\mcP_\zeta^{\rm obs}}\frac{\Lambda^2}{\Mpl^2}\right],
\label{bound NG}
\end{equation}
where $\mcP_\zeta^{\rm obs}\simeq 2.2\times10^{-9}$.
In the case of the parameters given in eq.~\eqref{ex para}, this upper bound leads to $N_G< 22.3$. 

We should also require that the energy density of the spectator scalar field is subdominant. Its energy fraction is given by
\begin{equation}
\Omega_\sigma = \frac{V(\bar{\sigma})}{3\Mpl^2 H^2} \simeq n \frac{\bar{\sigma} \Lambda}{\Mpl^2}.
\label{sigma fraction}
\end{equation}
Remembering $\dot{\bar \sigma}=-n\Lambda H$ during the growing phase and $\dot{\bar \sigma}=-\Lambda H$ during the attractor phase which terminates at $\bar{\sigma}\simeq \Lambda$,
the field value of $\sigma$ can be estimated as
\begin{equation}
\bar{\sigma}(t_k)\simeq \left(nN_G+N_A+1\right)\Lambda.
\end{equation}
Plugging it into eq.~\eqref{sigma fraction}, 
we obtain a constraint on the parameters,
\begin{equation}
\Omega_\sigma(t_k) \simeq n\left(nN_G+N_A+1\right)
\frac{\Lambda^2}{\Mpl^2} \ll 1.
\end{equation}
In the case of the parameters given in eq.~\eqref{ex para}, 
$\Omega_\sigma(t_k)\simeq 5\times 10^{-3}$ and the energy density of the spectator sector is subdominant.

 Finally, we put a constraint on the evolution of the two-form field after inflation.
 The background  form field decays as $a^{-2}$ which is slower than the radiation or matter components, and it might become dominant after the inflation.
 Defining the energy fraction of the background form field $\Omega_B \equiv \bar{\rho}_B/\rho_{\text{tot}}$, its expression is given by
\begin{equation}
\Omega_B \simeq \dfrac{\Lambda^2}{2\Mpl^2}e^{-2N_D}\left( \dfrac{a(t)}{a(t_{\text{end}})} \right)^2, 
\label{eq: ene}
\end{equation}
where $N_D$ is the duration of the number of e-foldings during the damping phase and $t_{\text{end}}$ is the time when inflation ends.
 Here, we assume an instant reheating so that our universe becomes radiation-dominated right after the end of inflation.
 Next, we estimate the amplitude of the curvature perturbation sourced by the form field perturbation after inflation.
 For an uniform-density slice, curvature perturbation is defined as $\zeta(t, \bm{x}) \equiv \ln[a(t, \bm{x})/a(t)]$ and its time evolution is given by
\begin{equation}
\dot{\zeta} = -\dfrac{H}{\rho_{\text{tot}} + p_{\text{tot}}}\delta p_{\text{nad}} \ , \label{eq: zetadot}
\end{equation}
where $\delta p_{\text{nad}} \equiv \delta p - \tfrac{\dot{\bar{p}}_{\text{tot}}}{\dot{\bar{\rho}}_{\text{tot}}}\delta \rho$ is the non-adiabatic pressure.
 Since $\rho_{\text{tot}}$ and $p_{\text{tot}}$ are the same order, on superhorizon the integration of \eqref{eq: zetadot} is approximately given by
\begin{equation}
\zeta_B(t,k) \sim \int^{a(t)}\dfrac{da}{a}\dfrac{\delta\rho_B}{\rho_{\text{tot}}} \simeq \dfrac{\delta\rho_B(t, k)}{\rho_{\text{tot}}(t)} \ ,
\end{equation}
where we used the fact that the integrand of the $a$ integral is an increasing function at the radiation-dominated era, because $\delta\rho_B$ on super-horizon is proportional to $a^{-2}$ which is a slower dilution than the energy density of radiation $\rho_{\text{tot}} \propto a^{-4}$.
 Hence, using the analytic expression for $\delta\rho_B(t,k)/\bar{\rho}_B(t)$ during the attractor phase, we obtain
\begin{equation}
\mathcal{P}_{\zeta_B}(k) \sim \Omega^2_B\dfrac{H_{\text{inf}}^2}{\Lambda^2}\left(\dfrac{k_A}{k}\right)^{2\Delta n} \ .
\end{equation}
 At CMB scales $k=k_{\text{CMB}}$, the condition $\mathcal{P}_{\zeta_B}(k_{\text{CMB}}) \ll 2.2\times10^{-9}$ leads to
\begin{equation}
\Omega_B \ll 1.7 \times 10^{-2} \label{eq: cur} \qquad \Longleftrightarrow \qquad \ln\left(\dfrac{a(t)}{a(t_{\text{end}})}\right) \ll N_D + 2.9
\end{equation}
in our set of parameters \eqref{ex para}.
This bound implies that the form field should become massive and decay into other particles within $N_D+2.9$ e-foldings after the inflation end.

\section{Conclusion}
\label{Conclusion}

 In this study, we developed the phenomenology of anisotropic inflation with two-form field.
 More precisely, we studied the model of inflation where a two-form field is kinetically coupled to a spectator scalar field.
 As to the background dynamics of the spectator field, we considered the situation where it slowly rolls down at first and get stabilized at a certain time on its potential.
 Depending on the evolution of the form field, the background dynamics is separated into three phases: (i) growing phase, (ii) attractor phase and (iii) damping phase.
 During the growing phase, the energy density of background form field $\bar{\rho}_E$ is negligibly small but grows as $a^{2\Delta n}$ due to the time variation of kinetic function.
 Simultaneously, on superhorizon scales perturbation of form field $\bar{I}\delta B/a^2$ also amplifies as $a^{\Delta n}$ which sources that of spectator field $\delta\sigma$ growing up as $a^{2\Delta n}$.
 When the backreaction of $\bar{\rho}_E$ becomes significant, $\bar{\rho}_E$ get balanced to the kinetic energy of the spectator field and stays constant.
 At this attractor phase, $\bar{I}\delta B/a^2$ and $\delta\sigma$ also stop growing and get constant values whose ratio depends on the angle of wave number $\theta$.
 Finally, at the damping phase $\sigma$ starts to oscillate around the minimum of potential and $\bar{\rho}_E$ decays as $a^{-2}$.
 We solved above dynamics and derived the analytical expressions of background and perturbation both of which are confirmed through numerical calculations.

 The main prediction of this work is that the sourced $\delta\sigma$ generates the statistically anisotropic gravitational waves via the presence of the background form field.
 Interestingly, only one linear polarization mode couples to the scalar perturbation and the resultant power spectrum is linearly polarized.
 This feature is distinct from another inflationary models with gauge fields topologically coupled to scalar sectors  \cite{Sorbo:2011rz, Cook:2011hg, Anber:2012du, Barnaby:2012xt, Dimastrogiovanni:2012ew, Adshead:2013qp, Mukohyama:2014gba, Obata:2014loa, Namba:2015gja, Obata:2016tmo, Domcke:2016bkh, Maleknejad:2016qjz, Guzzetti:2016mkm, Obata:2016xcr, Dimastrogiovanni:2016fuu, Adshead:2016omu, Garcia-Bellido:2016dkw, Obata:2016oym, Fujita:2017jwq, Thorne:2017jft, Ozsoy:2017blg, Agrawal:2018mrg}.
 Furthermore, we found that the resultant tensor power spectrum is written by the combination of angular functions $\cos^n\theta \ (n = 2, 4, 6)$ and the statistical anisotropy does not depend on any model parameters.
 Remarkably, the quadrupole moment vanishes at leading order and only higher harmonics appear.
 This result should be compared with the case of U(1) gauge field in our previous work \cite{Fujita:2018zbr} and can be an unique property from the phenomenology of inflation with two-form field.
 We estimated the detectability of the sourced gravitational waves.
 We derived several constraints on the parameters and show a viable example of parameter set where the amplitude of tensor power spectrum is detectable in near future.
 Since we have some concrete upcoming experiment such as LiteBIRD, it would be interesting to estimate the testable amplitude of the statistical anisotropy based on their realistic sensitivities~\cite{Hiramatsu:2018vfw}.
 We leave this issues for future work.

\section{Acknowledgement}

  In this work, TF is supported by Grant-in-Aid for JSPS Research Fellow No.~17J09103.

\appendix

\section{Gauge transformation of two-form field}
\label{Gauge transformation of form field}

 The action of form field is invariant under the following transformation
\begin{equation}
\delta B_{\mu\nu} = \partial_\mu \xi_\nu - \partial_\nu \xi_\mu \ .
\end{equation}
 Note that the parameter $\xi_\mu$ can be reduced to $\partial_i \xi_i = 0$ since it has the redundancy $\xi_\mu \rightarrow \xi_\mu + \partial_\mu \chi$.
 By using this degrees of freedom, we can choose
\begin{align}
&\partial_i B_{ij} = 0 \qquad (\text{choosing} ~\triangle\xi_i = -\partial_i B_{ij}) , \\
&\partial_i B_{i0} = 0 \qquad (\text{choosing} ~\triangle\xi_0 = -\partial_i B_{i0}) .
\end{align}
 Due to above transverse conditions, as to the non-dynamical valuable $\delta B_{0i}$ we can decompose it with the linear polarization vectors $e^X_i(\hat{\bm k})$ and $e^Y_i(\hat{\bm k})$ in Fourier space as
\begin{align}
\delta B_{0i}(t,\bm x)= 
\int \frac{\dd^3 k}{(2\pi)^3} e^{i\bm{k}\cdot\bm{x}}\, \left[e^X_i(\hat{\bm k}) \delta B_{\bm k}^X(t)+i e^Y_i(\hat{\bm k}) \delta B_{\bm k}^Y(t)\right]
\end{align}
and integrated out from the quadratic action.
 One can obtain the following constraint equations
\begin{align}
k\bar{I}\delta B^X_{\bm{k}} &= -\dfrac{\bar{I}\dot{\bar{B}}_{xy}\sin\theta}{\sqrt{2}}~h^\times_{\bm{k}} \label{eq: non1} \ , \\
k\bar{I}\delta B^Y_{\bm{k}} &= \bar{I}\dot{\bar{B}}_{xy}\sin\theta\left( 2\dfrac{\bar{I}'}{\bar{I}}\delta\sigma_{\bm{k}} + \dfrac{h^+_{\bm{k}}}{\sqrt{2}} - \dfrac{1}{2\Mpl^2H}\left(  \dot{\bar{\phi}}\delta\phi_{\bm{k}} + \dot{\bar{\sigma}}\delta\sigma_{\bm{k}} + \dfrac{\bar{I}^2\dot{\bar{B}}_{xy}}{a^4}\delta B_{\bm{k}} (i\cos\theta) \right) \right) \label{eq: non2} \ .
\end{align}

\section{Quadratic Action}
\label{Quadratic Action}

Here we show the reduced expression of the second order action
of $\dphi, \dsig, \delta B_{ij}$ and $h_{ij}$.
Integrating the non-dynamical component of scalar metric perturbations such as lapse and shift function and temporal component of two-form field, we get
\begin{equation}
S^{(2)}=\frac{1}{2}\int \dd t\dd^3 x\, a^3  \Big[ \mathcal{L}_{\rm scalar}
+\mathcal{L}_{\rm form} + \mathcal{L}_{\rm tensor} \Big].
\end{equation}
The Lagrangians of the scalar, two-form and tensor sectors are given by 
\begin{align}
\mathcal{L}_{\rm scalar}&= 
\dphid^2+\dsigd^2-a^{-2}(\partial_i \dphi)^2-a^{-2} (\partial_i \dsig)^2 -\mu_\phi^2 \dphi^2-\mu_\sigma^2\dsig^2-2\Omega_{\phi\sigma}\dphi\dsig,
\\
\mathcal{L}_{\rm form}&= 
\frac{\bar{I}^2}{2a^4}\bigg[\delta \dot{B}_{ij}^2-a^{-2}(\partial_i\delta B_{jk})^2
-\mu^{2}_{ijkl}\delta B_{ij} \delta B_{kl} -2 \Omega^{B\phi}_{ij}  \delta B_{ij} \dphi
- 2\Omega^{B\sigma}_{ij}  \delta B_{ij} \dsig
+4\frac{\bar{I}'}{\bar{I}} \dot{\bar{B}}_{ij}\delta\dot{B}_{ij}\dsig
\bigg], \\
\mathcal{L}_{\rm tensor}&= \dfrac{\Mpl^2}{4}\left( \dot{h}_{ij}\dot{h}_{ij} - a^{-2}\partial_k h_{ij}\partial_k h_{ij} \right) \notag \\
& + \dfrac{\bar{I}^2}{2a^4}\left[ \dot{\bar{B}}_{ij}\dot{\bar{B}}_{kl}h_{ik}h_{jl} - \dfrac{2}{3} \dot{\bar{B}}_{ij}\dot{\bar{B}}_{ij}h_{kl}h_{kl} + 4\dot{\bar{B}}_{ij}\delta\dot{B}_{jk}h_{ik} + 2\dfrac{\bar{I}'}{\bar{I}}\dot{\bar{B}}_{ij}\dot{\bar{B}}_{jk}h_{ik}\delta\sigma \right] \notag \\
& + \dfrac{\bar{I}^2\dot{\bar{B}}_{ij}\dot{\bar{B}}_{jk}}{4a^4\Mpl^2H}h_{ik}\left( \dot{\bar{\phi}}\delta\phi + \dot{\bar{\sigma}}\delta\sigma + \dfrac{\bar{I}^2}{2a^4}\dot{\bar{B}}_{ij}\delta B_{ij} \right) \label{eq: GWq}
\end{align}
with
\begin{align}
\mu_\phi^2&\equiv 
\bar{U}''
-3\frac{\dot{\bar \phi}^2}{\Mpl^2}\left(1+\frac{\epsilon_H}{6}+\frac{2\ddot{\bar \phi}}{3H\dot{\bar \phi}}+\frac{\dot{\bar \phi}^2 +\dot{\bar \sigma}^2
+2\bar{\rho}_E \cos^2\theta}{12\Mpl^2 H^2}\right),
\\
\mu_\sigma^2&\equiv 
\bar{V}''
+\frac{\dot{\bar{B}}_{ij}^2}{2a^4}\left(4{\bar I}'^2 \sin^2\theta -{\bar I}'^2-\bar{I} \bar{I}''\right)
\notag\\
&\quad-3\frac{\dot{\bar \sigma}^2}{\Mpl^2}\left(1
-\frac{\bar{I} \bar{I}' \dot{\bar{B}}_{ij}^2}{3a^4 H\dot{\bar \sigma}}\cos^2\theta+\frac{\epsilon_H}{6}+\frac{2\ddot{\bar \sigma}}{3H\dot{\sigma}}+\frac{\dot{\bar \phi}^2 +\dot{\bar \sigma}^2
+2\bar{\rho}_E \cos^2\theta}{12\Mpl^2 H^2}\right),
\\
\Omega_{\phi\sigma} &\equiv 
-\frac{\dot{\bar \phi}\dot{\bar \sigma}}{\Mpl^2}\left[3
-\frac{2\bar{\rho}_E\cos^2\theta}{\Lambda H\dot{\bar \sigma}}+\frac{\epsilon_H}{2}+
\frac{\dot{\bar \phi}^2 +\dot{\bar \sigma}^2
+2\bar{\rho}_E\cos^2\theta}{4\Mpl^2 H^2}+\frac{\ddot{\bar \phi}}{H\dot{\bar \phi}}+\frac{\ddot{\bar \sigma}}{H\dot{\bar \sigma}}\right],
\\
\mu^{2}_{ijkl}&\equiv \frac{3\bar{I}_B^2 \dot{\bar{B}}_{ij} \dot{\bar{B}}_{kl}}{2a^4\Mpl^2}\left(1 + \dfrac{2}{3}\tan^2\theta - \frac{\epsilon_H}{6}-\frac{\dot{\bar \phi}^2+\dot{\bar \sigma}^2
+2\bar{\rho}_E\cos^2\theta}{12\Mpl^2H^2}\right),
\\
\Omega^{B\phi}_{ij} &\equiv -\frac{\dot{\bar \phi}\dot{\bar{B}}_{ij}}{2\Mpl^2}\left(\epsilon_H+\frac{2\ddot{\bar \phi}}{H\dot{\bar \phi}}+\frac{\dot{\bar \phi}^2+\dot{\bar \sigma}^2
+2\bar{\rho}_E \cos^2\theta}{2\Mpl^2 H^2}\right),
\\
\Omega^{B\sigma}_{ij} &\equiv  \frac{\dot{\bar{B}}_{ij}}{4\Mpl^2}\left[\frac{\bar{I} \bar{I}' \dot{\bar{B}}_{ij}^2}{a^4 H}\cos^2\theta-\dot{\bar \sigma}\left(\epsilon_H+\frac{2\ddot{\bar \sigma}}{H\dot{\bar \sigma}}+\frac{\dot{\bar \phi}^2+\dot{\bar \sigma}^2+2\bar{\rho}_E \cos^2\theta}{2\Mpl^2 H^2}\right)\right],
\end{align}
where $\epsilon_H\equiv -\dot{H}/H^2$.



\end{document}